\documentclass[12pt]{article}
\usepackage{amssymb,amsmath,amsthm}
\usepackage{array,longtable}
\usepackage[dvips]{graphicx}
\usepackage[dvips]{color}
\usepackage{wrapfig}
\usepackage{epsfig}
\usepackage[mathscr]{eucal}

\newcommand{\Cc}{\mathcal{C}}
 \newcommand{\pd}{\partial}
  \newcommand{\ap}{\alpha^{\prime}}

\begin{document}
\title{
%$~$\\
\Large{ \textup{Time Evolution in Superstring Field Theory on
non-BPS brane. \\}} \Large{ \textup{I. Rolling Tachyon
and Energy-Momentum Conservation
\\}} \Huge{ \textup{  }}}

\author{
I.~Ya.~Aref'eva$^1$, L.~V.~Joukovskaya$^2$ and A.~S.~Koshelev$^{1,3}$
\vspace{5mm}\\
$^1$Steklov Mathematical Institute\\
Russian Academy of Sciences\\
Gubkin st. 8, Moscow, 119991, Russia
\vspace{1mm}
\\
\textsf{arefeva@mi.ras.ru~~~alex-kas@yandex.ru}
\vspace{6mm}\\
$^2$Physics Department\\
Moscow State University\\
Moscow, 119899, Russia
\vspace{1mm}\\
\textsf{l$_-$joukovskaya@mtu-net.ru}
\vspace{5mm}\\
$^3$Physics Department\\
University of Crete, Greece\\
}

\date{~}
\maketitle \thispagestyle{empty}

\begin{abstract}
We derive  equations of motion  for the tachyon field living on an
unstable non-BPS D-brane in  the level truncated  open cubic
superstring field theory in the first non-trivial approximation.
We construct a special time dependent solution to this equation
which describes the rolling tachyon.
It starts from the perturbative vacuum and  approaches   one of
stable vacua  in infinite time.
We investigate conserved energy functional and show that its different parts
dominate in different  stages of the evolution.
We show that the
pressure for this solution has its minimum at zero time and
goes to minus energy at infinite time.

\end{abstract}

\newpage
\tableofcontents
\section{Introduction}
The physical problem which we would like to address in a series of
papers   is a time dependence of the tension of an unstable
non-BPS D-brane in the process of its decay. It is  expected that
starting from its maximal value, equal to the density of the
vacuum energy, the tension of an unstable non-BPS D-brane becomes
zero  when the tachyon field approaches the true vacuum. In this
paper we start with  study of  classical time dependent solutions
describing a motion of the tachyon on an unstable non-BPS D-brane
within level truncation scheme in cubic open superstring field
theory.  In the next paper  we will  take into account  higher
level fields including vector fields and  study a dynamics of a
D-brane tension.

The  tachyon potential and various classical solutions in SFT on
an unstable D-brane system have been intensively studied in the
last two years (see \cite{Sen,Sen2,0005036,BSZ,TM}, reviews
\cite{0102085,ABGKM,0301094} and references therein). In the
beginning only time independent solutions representing either the
tachyon vacuum or static D-branes of lower dimension have been
examined. The process of production and decay of unstable branes
can be  described  as a space-like D-brane \cite{0202210}.
For earlier study of the tachyon dynamics on brane-antibrane
system see, for example, \cite{older}.

Recently, Sen has proposed to study the classical decay process of
an unstable D-brane in the framework of SFT
\cite{0203211,0203265,0204143,0207105}. As compare to the usual
local field theory the time dependent equation of  SFT containing a
tachyon has  special features. First, this equation involves
infinite number of other fields. Second,  the string field theory
action contains infinite number of time derivatives, and hence the
initial value problem seems to be not well-defined. Nevertheless,
Sen \cite{0203211,0207105} has shown that it is possible to
construct a family of classical solutions to the string field
theory equations of motion characterized by the initial position
and velocity of the tachyon field.

There are at least two different ways to construct time dependent
solutions within SFT:
\begin{itemize}
\item
One can study  solutions to  SFT equation of motion
\begin{equation}
\label{SFTEOM} QA+A*A=0
\end{equation}
perturbatively, specifying some  initial data.  This way has
been followed in the background approach to scattering theory in
SFT in \cite{AMZ-PL} as well as in  Sen's paper \cite{0207105}.
\item
One can also study a tachyon dynamics within the level truncation
scheme. As it has been noted in \cite{0207107} there are many
similarities  between the tachyon equation of motion obtained in
SFT via the level truncation and the p-adic string equation of
motion \cite{VVZ,FO,BFOW}.
\end{itemize}

The goal of this paper is to study the  tachyon dynamics in SSFT
on a non-BPS D-brane  within the level truncation scheme
\cite{ABKM}. We derive the equation of motion for the tachyon in
the first nontrivial approximation. We construct a special time
dependent solution to this equation which starts from the
perturbative vacuum and approaches  one of stable vacua.  To
construct this solution we consider it as  a positive time part
of a solution  that  interpolates  between two non-perturbative vacua.
 As in the case of the p-adic string \cite{BFOW} to
perform numerical study of the SSFT tachyon equation of motion
with boundary conditions on $\pm $ infinities, it
is useful to deal with the integral form of the equation of motion.
 The integral form of the
tachyon equation of motion in SSFT is similar to the integral form
of the tachyon equation of motion in the p-adic string \cite{BFOW}
for $p=3$ while the integral form of the bosonic SFT   is similar
to one of the p-adic string for $p=2$ \cite{0207107,Yang}. In
contrast to p-adic case, kernels of the integral equations
describing SFT and SSFT tachyons are not positively defined and
application of an iteration procedure is more subtle. By our
request these simulations have been performed in \cite{yar} and we will
use these results in the paper. Numerical analysis \cite{yar} is
based on an iteration procedure of solving nonlinear integral
equations. It is interesting to note  that SFT and SSFT tachyons
equations can be considered as perturbations of the  corresponding
p-adic string equations of motion. The superstring interpolating
tachyon solution can be obtained in a linear approximation around
the corresponding p-adic (p=3) solution. This approximation
rather well reproduces results of numerical calculations.

In order to understand qualitative  properties of solutions we write down
the conserved energy functional. The energy $E$ is the sum of
three terms, namely  the kinetic term $E_k$, the
potential term $E_p$ and the non-local term $E_{nl}$. There
are no difficulties  to study numerically the kinetic $E_k$ and
potential $E_p$ terms. To investigate numerically contribution of
the non-local term $E_{nl}$ we write it in a special form
admitting an integral representation.

The pressure is also the sum of three terms
\begin{equation}
\label{presu}
P(t)=-E+2E_k(t)+E_{nlp}(t).
\end{equation}
The non-local term $E_{nlp}$ is just a part of the  non-local term
$E_{nl}$ contributing to the energy $E$. We show that for large
time $P(\infty)=-E$ while for $t=0$, under a special approximation
for the tachyon solution to the SSFT equations of motion (see Sect.~2.2),
$P(0)\approx E$. The energy $E$ is equal to the tachyon potential minimum and according
to Sen's
conjecture the tension of the non-BPS D-brane compensates this energy and the total
energy of a system is equal to zero.
This compensation has been checked for cubic SSFT with rather good accuracy
\cite{ABKM}. Since the total energy is equal to zero the total pressure also
vanishes at large time for our solution.
 We also show that the
pressure for this solution has its minimum at zero time and
goes to minus energy at infinite time.

Let us also note that there are other approaches to study
rolling tachyons. One of them is related to  CFT description of a
non-BPS D-brane dynamics. Sen
 has noted \cite{0203211,0203265,0207105} (and refs. therein) that
 dynamics of a tachyon on an unstable D-brane
can be described by an effective  Born-Infeld action with an exponentially
decreasing potential. Though this action cannot be  derived from
first principles it has some  features  desirable in  a
classical dynamics of open SFT around the tachyon vacuum. Namely,
in this effective field theory there are solutions with
fixed energy density and asymptotically vanishing pressure. These
solutions are also interesting from cosmological point of view, see
\cite{0209122} for details.

Let us note that it would be interesting to see more relations
between SFT and the effective  Born-Infeld action. An incorporation of
vector field could  clarify this problem. This hope is supported
by  the observation that an incorporation of vector fields, apparently,
reproduces the Born-Infeld action \cite{0001201,0003221,effective}.

The paper is organized as follows.
In Section~2 we introduce a general setup, namely the cubic
superstring field theory action for a description of an open
superstring living on a non-BPS D-brane. Within the level
truncation scheme
we get from this action tachyon equations of motion. In fact in this paper we
restrict ourself to  the first
non-trivial approximation which includes the tachyon field and one
auxiliary field. Under an assumption that even
derivatives of the auxiliary field vary slowly we
get an approximate version of the tachyon equation of motion.
Then we  rewrite differential equations in the integral form.
In Section~3 we present results of numeric study of the equations
obtained in Section~2.

In Section~4 functionals of the energy and pressure are studied.
Conservation of energy is calculated explicitly. For the
pressure it is shown that it has minimum at zero time and equal
to minus energy at infinite time.

\section{Equations of Motion for Tachyon Field on Non-BPS D-brane}

To describe the open string states living on a single non-BPS
$\mathrm{D}$-brane one has to consider GSO$\pm$ states \cite{Sen}.
GSO$-$ states are Grassmann even, while GSO$+$ states are
Grassmann odd. The unique (up to rescaling of the fields) gauge
invariant cubic action unifying GSO$+$ and GSO$-$ sectors is
\cite{ABKM}
\begin{equation}
\begin{split}
S[A_+,A_-]&=\frac{1}{g^2_o}\left[ \frac{1}{2}\langle\!\langle
Y_{-2}|A_+,Q_BA_+ \rangle\!\rangle+\frac{1}{3}\langle\!\langle
Y_{-2}|A_+,A_+,A_+\rangle\!\rangle
\right.\\
&~~~~~~~~~\left.+\frac{1}{2}\langle\!\langle
Y_{-2}|A_-,Q_BA_-\rangle\!\rangle -\langle\!\langle
Y_{-2}|A_+,A_-,A_-\rangle\!\rangle\right].
\end{split}
\label{action7}
\end{equation}
For $A_-=0$ one gets the  action \cite{AMZ-np,PTY}. Here the
factors before the odd brackets are fixed by the constraint of
gauge invariance, that is specified below, and the reality of the
string fields $A_{\pm}$. Variation of this action with respect to
$A_+$ and $A_-$ yields the following equations of motion (see
\cite{ABKM} for notations and details)
\begin{equation}
Q_BA_{\pm}+A_+\star A_{\pm} -A_-\star A_{\mp}=0.
 \label{eqmotion}
\end{equation}
The action \eqref{action7} is invariant under the gauge
transformations
\begin{equation*}
\delta A_{\pm}=Q_B\Lambda_{\pm}+[A_{\pm},\Lambda_+],
+\{A_{\mp},\Lambda_-\},
\end{equation*}
where $[\,,]$ ($\{\,,\}$) denotes $\star$-(anti)commutator and
$\Lambda_{\pm}$ are gauge parameters.

To perform actual calculations the level truncation scheme is
employed.

Let us denote by $\mathcal{H}_1$ the subset of vertex operators of
ghost number $1$ and picture $0$, created by the matter stress
tensor $T_B$, matter supercurrent $T_F$ and the ghost fields $b$,
$c$, $\pd\xi$, $\eta$ and $\phi$. We restrict the string fields
$\mathcal{A}_+$ and $\mathcal{A}_-$ to be in this subspace
$\mathcal{H}_1$.

Next we expand $\mathcal{A}_{\pm}$ in a basis of $L_0$
eigenstates, and write the action (\ref{action7}) in terms of
space-time component fields. The string field is now a series with
each term being a vertex operator from $\mathcal{H}_1$ multiplied
by a space-time component field. We define the level $K$ of string
field's component $A_i$ to be $h+1$, where $h$ is the conformal
dimension of the vertex operator multiplied by $A_i$, i.e. by this
convention the tachyon is taken to have level $1/2$. To compose
the action truncated at level $(K,\,L)$ we select all the
quadratic and cubic terms  of total level not more than $L$ for
the space-time fields of levels not more than $K$. Since our
action is cubic, $2K\le L \le 3K$.

\subsection{Action and equations of motion}

The level (1/2,1) action contains the tachyon field $\phi$ and one
auxiliary field $u$
\begin{equation}
S[u,\phi]=\frac{1}{g_o^2\alpha^{\,\prime \frac{(p+1)}{2}}} \int\;
d^{p+1}x\left[ u^2(x)- \frac{\ap}2
\partial _\mu \phi(x)\partial ^\mu \phi(x)+\frac14 \phi^2(x) -
\frac{1}{3\gamma^2} \tilde{u}(x)\tilde{\phi}(x)\tilde{\phi}(x)
\right], \label{action-x}
\end{equation}
where the tilde  is the  action of the operator
\begin{equation}
\label{tilde-u} D_\Box =
\exp(-\alpha^{\,\prime}\ln\gamma\,\Box),~~~\tilde{\phi}=D_\Box
\phi, ~~~{\mbox{etc}}.
\end{equation}
$\gamma=\frac4{3\sqrt{3}}$,  $\Box=\pd_{\mu}\pd^{\mu}$ and the
signature  is $(-+++\dots)$. The equations of motion have the form
\begin{subequations}
\label{up}
\begin{equation}
\label{seom-u} 2u(x) -\frac{1}{3\gamma^2}D_\Box\left((D_\Box
\phi)^2\right)(x)=0,
\end{equation}
\begin{equation}
\label{seom-t} \left(\alpha^{\,\prime}\Box+1/2\right)\phi(x)
-\frac{2}{3\gamma^2}D_\Box\left((D_\Box \phi) (D_\Box
u)\right)(x)=0.
\end{equation}
\end{subequations}
Applying $D_\Box$ to both equations above we can write them as
\begin{subequations}
\label{PhiPsi}
\begin{equation}
\label{Phi} \tilde{u}
-\frac{1}{6\gamma^2}D^2_\Box\left((\tilde{\phi})^2\right)=0,
\end{equation}
\begin{equation}
\label{Psi} \left(\alpha^{\,\prime}\Box+1/2\right)\tilde{\phi}
-\frac{2}{3\gamma^2}D^2_\Box\left( \tilde{u}\tilde{\phi}\right)=0.
\end{equation}
\end{subequations}
Here $D^2_\Box \phi \equiv
\exp(-2\alpha^{\,\prime}\ln\gamma\,\Box)\phi$ and for simplicity
we do not write arguments of functions. Performing a rescaling of
fields and coordinates
\begin{equation}
\Psi(x)=\frac{\sqrt{2}}{3(\gamma)^2}\tilde{\phi}(\alpha
x)~~\text{and}~~ \Upsilon(x)=\frac4{3\gamma^2}\tilde{u}(\alpha x)
\label{psi-up-m}
\end{equation}
with
\begin{equation}
\alpha^2=4a=-4\alpha'\ln \gamma  ~~~\mbox {and}~~~
q^2=\frac{\alpha'}{2a}=-\frac{1}{4\ln \gamma}.
\end{equation}
We recast the latter system to the following form
\begin{subequations}
\label{psi-up-m2}
\begin{equation}
\Upsilon(x) -D^2_\Box\left(\Psi^2\right)(x)=0,
\end{equation}
\begin{equation}
\left(q^2\Box+1\right)\Psi(x) -D^2_\Box\left( \Upsilon
\Psi\right)(x)=0.
\end{equation}
\end{subequations}
 Hereafter $D_\Box=\exp(\frac18 \Box)$.

\subsection{Approximation $u=\tilde{u}$}

The approximation $u=\tilde{u}$  is valid  when  even derivatives
of the field $u$  vary slowly. Substituting $u=\tilde{u}$ in
action (\ref{action-x}) we are left with the action which after
integration over $u$ and rescaling has the  form
\begin{equation}
S[\psi]= \int\; dx\left[\frac12 \psi^2(x)
-\frac{q^2}{2}\partial_\mu \psi(x)\partial^\mu \psi(x) -
\frac14\tilde{\psi}^4(x) \right], \label{action-tab-x}
\end{equation}
where
\begin{equation}
\label{tilde-psi} \tilde{\psi}(x)\equiv\Psi (x)= \exp(\frac18
\Box)\psi(x).
\end{equation}
Equation of motion for this action is
\begin{eqnarray}
\label{Psi-ap} \left(q^2\Box+1\right)\Psi(x) =D^2_\Box\left( \Psi
^3\right)(x),
\end{eqnarray}
 which can be presented as a system of two equations
\begin{subequations}
\label{psi-up-ap}
\begin{equation}
\Upsilon(x) -\Psi^2(x)=0,
\end{equation}
\begin{equation}
\left(q^2\Box+1\right)\Psi(x) -D^2_\Box\left( \Upsilon
\Psi\right)(x)=0.
\end{equation}
\end{subequations}

\subsection{Integral equations for space homogeneous configurations }

For space homogeneous configurations one has
\begin{subequations}
\label{Phsi1}
\begin{equation}
\label{Phi1} \Upsilon(t) -D^2_t\left(\Psi^2\right)(t)=0,
\end{equation}
\begin{equation}
\label{Psi1} \left(-q^2\frac{d^2}{dt^2}+1\right)\Psi(t)
-D^2_t\left( \Upsilon \Psi\right)(t)=0.
\end{equation}
\end{subequations}
Initial  and tilded  fields  are related as
\begin{subequations}
\begin{equation}
\label{tilde-psi1} \tilde{\psi}(t)\equiv\Psi (t)= e^{-\frac18
\frac{d^2}{dt^2}} \psi(t), ~~~~
\end{equation}
\begin{equation}
\label{tilde-phi1} \tilde{\upsilon}(t)\equiv\Upsilon (t)=
e^{-\frac18 \frac{d^2}{dt^2}} \upsilon(t). ~~~~
\end{equation}
\end{subequations}
Note that the  operator $D^2_t=e^{-\frac14\frac{d^2}{dt^2}}$ is
not well-defined  on an arbitrary  smooth function. This happens
because the operator in the exponent is positively defined.
However, we can rewrite the  system of equations (\ref{Phsi1})
 using only the
operator $D^{-2}_{t}=e^{\frac14\frac{d^2}{dt^2}}$ . The resulting
system is
\begin{subequations}
\label{18}
\begin{equation}
\label{Phi-m} D^{-2}_{t}\Upsilon =\Psi^2,
\end{equation}
\begin{equation}
\label{Psi-m}
 \left(-q^2
\frac{d^2}{dt^2}+1\right)D^{-2}_{t}\Psi = \Upsilon \Psi.
\end{equation}
\end{subequations}
As in the case of p-adic string \cite{0207107} one can rewrite
this system of differential equations in the integral form
\begin{subequations}
\label{seq}
\begin{equation}
\label{Phi-con}
 C \Upsilon=\Psi^2,
\end{equation}
\begin{equation}
\label{Psi-con}
 (- q^2 \frac{d^2}{dt^2}+1)C\Psi=\Upsilon \Psi,
\end{equation}
\end{subequations}
where
$$
\Cc \Psi \equiv \Cc_{1/4}\Psi
$$
and
\begin{equation}
\label{con-def} \Cc_a\psi (t)=\frac{1}{\sqrt{4\pi a}}\int
_{-\infty}^{\infty} e^{-\frac{(t-t')^{2}}{4a}} \psi(t')dt'.
\end{equation}
The initial fields are related with tilded  ones by means of the
integral operator $C_a$
\begin{subequations}
\label{til}
\begin{equation}
\label{up-Up}
\upsilon=\Cc_{1/8}\Upsilon=\Cc_{1/8}\tilde{\upsilon},
\end{equation}
\begin{equation}
\psi=\Cc_{1/8}\Psi=\Cc_{1/8}\tilde{\psi}. \label{psi-Psi}
\end{equation}
\end{subequations}
In the case $q^2=0$ equations (\ref{18}) take a simple form
\begin{subequations}
\label{Phsi-con-sl}
\begin{equation}
\label{Phi-con-sl}
 C \Upsilon=\Psi^2,
\end{equation}
\begin{equation}
\label{Psi-con-sl} C\Psi=\Upsilon \Psi.
\end{equation}
\end{subequations}

\subsection{Approximate integral equation}

On the space homogeneous configurations instead of (\ref{Psi-ap})
one gets
\begin{eqnarray}
\label{Psi-ap1} \left(-q^2\frac{d^2}{dt^2}+1\right)D^{-2}_t\Psi
=\Psi ^3.
\end{eqnarray}
This equation can be rewritten in the integral form
\begin{eqnarray}
\label{Psi-ap2} \left(-q^2\frac{d^2}{dt^2}+1\right)\Cc\Psi =\Psi
^3,
\end{eqnarray}
where $\Cc\Psi \equiv \Cc_{1/4}\Psi$ and $\Cc_{1/4}$ is defined
in (\ref{con-def}).
 If we drop the term with the second
derivative we are left with
\begin{eqnarray}
\label{p=3} \Cc\Psi =\Psi ^p,~~~p=3.
\end{eqnarray}
This is a p-adic effective field equation for p=3
\cite{VVZ,BFOW,0207107}. For the bosonic string theory one gets
equation (\ref{p=3}) with p=2. As it has been shown in
\cite{0207107} properties of p-adic equation are drastically
different for p=2 and p=3. In particular, for p=2  there are no
rolling tachyon solutions.

\section{Solutions to Integral Equations}
\subsection{Interpolating solutions}

\begin{figure}[h!]
\centering
\includegraphics[width=6cm]{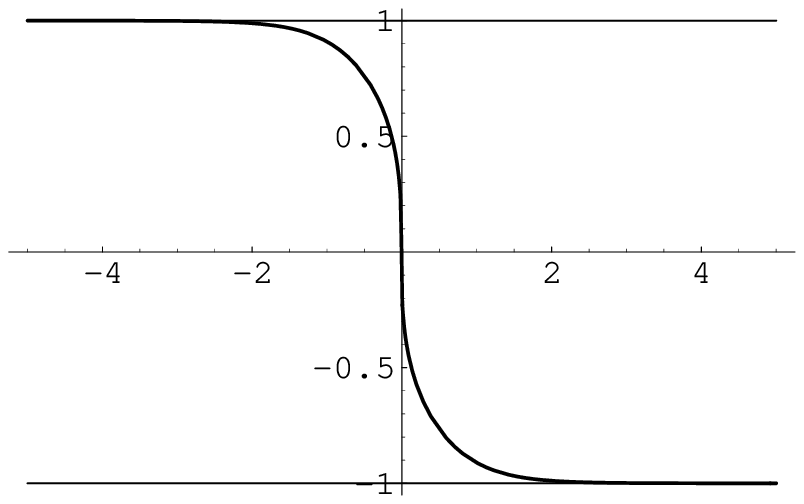}a$~~~$
\includegraphics[width=7cm]{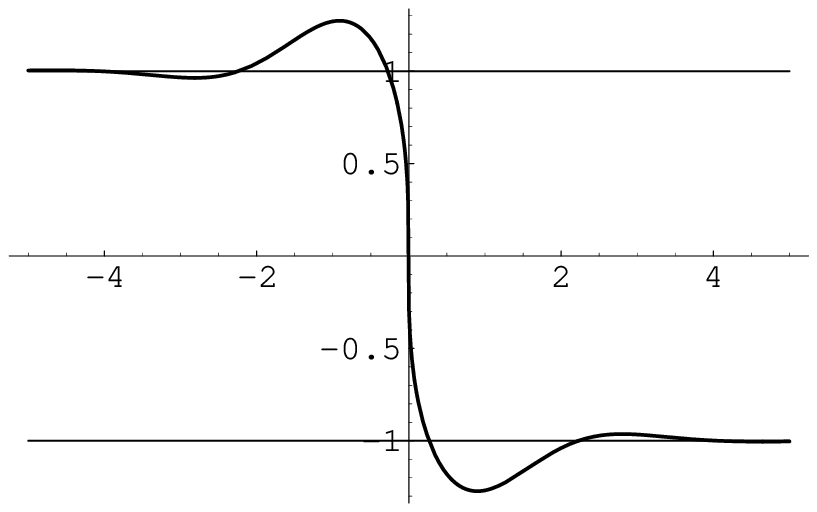}b
\caption{a)~$\Psi_0=\Psi_{app}$ for $q^2=0$; b)~$\Psi_{app}$ for
$q^2=0.96$.} \label{f12}
\end{figure}

\begin{figure}[h!]
\centering
\includegraphics[width=7cm]{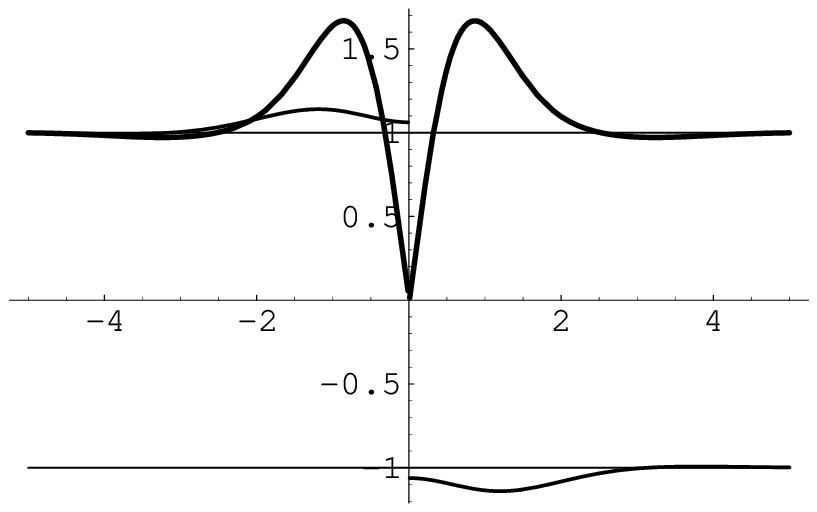}a $~~~$
\includegraphics[width=7cm]{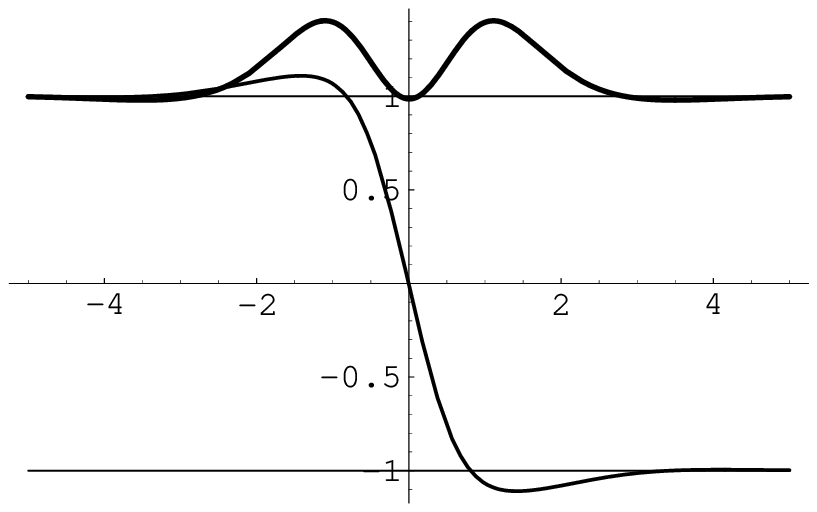}b
\caption { a) $\Psi$ (thin) and $\Upsilon $ (thick)  for
$q^2=0.96$;$~~~~~~~~~~~~~~~~~~~~~~~~~~~~~~~~~~~~~~~~~~~~~~~~~~~~~~~~~~~~~~$
 $~~~~~~~~~~~~~~$b) smoothed functions: $\psi $ (thin) and $\upsilon$ (thick) for $q^2=0.96$.}
\label{f3}
\end{figure}

\begin{figure}[h!]
\centering
\includegraphics[width=6.5cm]{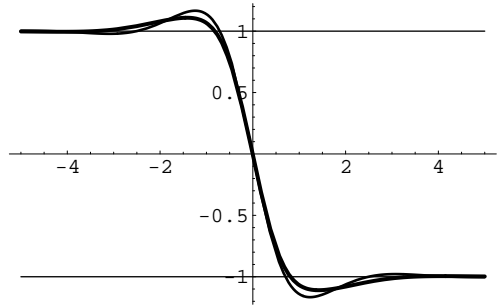}
\caption{Smoothed functions $\psi_{app}$ (thin) and $\psi$ (thick)
of  solutions for $q^2=0.96$.} \label{f5}
\end{figure}

Equation (\ref{p=3}) has been studied numerically in \cite{BFOW}
for p-adic effective field equations for arbitrary $p~$ and
recently in \cite{0207105}. More general equation (\ref{Psi-ap2})
as well as a system of equations (\ref{seq}) has been studied in
\cite{yar}. Here we review results of numerical calculations
performed in \cite{yar}. It was shown that:
\begin{itemize}
\item
Solution of (\ref{p=3}) interpolating between two vacua is given
by a monotonic function $\Psi_0$ presented in Fig.\ref{f12}a.
\item Solution of (\ref{Psi-ap2}) for $q^2\leq q_{cr}^2=1.38$ is
given by a  function $\Psi $  interpolating between two vacua
(see Fig.\ref{f12}b). This function oscillates near $\Psi_0$ with
exponentially decreasing amplitude.
\item
Function $\Psi$ solving the system (\ref{seq}) for $q^2\leq
q_{cr}^2=2.2$ is an odd function with a jump at $t=0$ and
$\Psi(\pm \infty)=\mp1.$ $\Upsilon$  is even function. In
Fig.\ref{f3}a these functions for $q^2\simeq 0.96$ are presented.
\item The smoothed functions $\upsilon$ and $\psi$, defined by (\ref{til}),
are continuous, see Fig.\ref{f3}b.
\item For $q^2\simeq 0.96$ function $\psi$ (\ref{psi-Psi})
does not differ much from the function obtained from the solution
of (\ref{Psi-ap2}), by the same formula (\ref{psi-Psi}) $\psi
_{app}=C_{1/8}\Psi _{app}$  (see Fig.\ref{f5}).
\end{itemize}

\subsection{Rolling solutions}

The  iterative procedure has been formulated so that the function
$\Psi$ is an odd one and the function $\Upsilon$ is even. Results
of calculations show that the function  $\Upsilon$ is smooth while
the function  $\Psi$
 has a discontinuity at the zero time. These results take place for $q \leq q_{cr},$
 including $q=0$.  For $q=0$ one can see that an assumption about the parity properties
  implies that $\Psi$ in fact has a discontinuity at $t=0$. Indeed,
  equations of motion (\ref{Phsi-con-sl})
can be rewritten as a system of equations  on semi-axis:
\begin{subequations}
\label{eq-on-sa}
\begin{equation}
\label{eq+} ( C_{+} \Upsilon)(t) =
\Psi(t)^2,~~~~~~~~~~~~~~~~~~~~~~
\end{equation}
\begin{equation}
\label{eq-} (C_{-} \Psi) (t)=\Upsilon (t)\Psi(t), ~~~0\leq
t<\infty,
\end{equation}
\end{subequations}
where $C_{+}$ and $C_{-}$ have kernels $K_{+}$ and $K_{-}$,
respectively,

\begin{subequations}
\label{Kpm}
\begin{equation}
 K_{+}(t,t') =K(t-t')+K(t+t'),
\end{equation}
\begin{equation}
K_{-}(t,t') =K(t-t')- K(t+t'),
\end{equation}
\end{subequations}
$K(t)=\frac{1}{\sqrt{\pi}}\exp (-t^2)$. These kernels are positive
when both $t,t'>0$. If one assumes that $0 < \Psi (t)< \infty $
for $0<t<\infty $ than it follows from the equation
\begin{equation}
\Upsilon =\frac{C_{-} \Psi}{\Psi},
\end{equation}
that $ \Upsilon (t)>0$ for  $0<t<\infty $. This guaranties  that
$(C \Upsilon)(t) >0$ for  $0\leq t<\infty $. According to
(\ref{eq+}) this means that $\Psi^2 >0$ for $t\geq 0$, i.e.
$\Psi^2 (0) >0.$ Since $\Psi$ is an odd function it has to have
discontinuity  at zero time.

This discontinuity disappears when we perform smoothing of this
function with the kernel $K$ to obtain the function $\psi$. In
fact this function represents the time evolution of the tachyon
field. Shown in thin line, for comparison is $\psi _{app}$ defined
as smoothing of the solution to  equation (\ref{Psi-ap2}).

We can consider a function  $\psi(t)$ starting  at zero time from
false  vacuum and rolling  to the true vacuum at large times. For
$q=0$ this is a monotonic function. As it was shown in \cite{yar}
for $q^2<q_{cr}^2$ there are small oscillations  with decreasing
amplitudes around the solutions to equation (\ref{p=3}).  The
presence of this oscillations (in fact oscillations with a complex
frequency) results from the presence of infinite number of
derivatives in a linear equation describing a deviation of
$\psi(t)$ from $\psi_{q=0}(t)$.

\section{Energy and Pressure}
\subsection{Conservation of energy}
In this section we derive functional expressions for energy for
tachyon equations (\ref{18}) and for approximate equation
(\ref{Psi-ap1}). Let us consider the system of integral equations
(\ref{seq}).
 The conserved energy is a sum of three terms representing kinetic energy $E_k$,
 potential energy $E_p$ and non-local terms
\begin{equation}
\label{parts-of-Energy}
 E(t)=E_{k}(t)+ E_{p}(t)+E_{nl}(t),
\end{equation}
where
\begin{equation}
\label{pk-of-Energy} E_{k}(t)=\frac{q^2}{2} (\partial \psi)^2,
\end{equation}

\begin{equation}
\label{pp-of-Energy}
 E_{p}(t)=- \frac{1}{4}\upsilon^2
-\frac12 \psi^2+  \frac{1}{ 2}(\Upsilon \Psi^2 ),
\end{equation}
\begin{eqnarray}
\label{pl-of-Energy} E_{nl}(t)= \frac18\int_0^1 d\rho\left(
e^{-\frac{\rho}{8}{\partial}^2}\Upsilon \Psi\right)
\overleftrightarrow{\partial}
  \left(\partial_{t}  e^{\frac{\rho}{8}{\partial}^2} \Psi \right)+
  \frac{1}{16} \int_0^1 d\rho\left(
e^{-\frac{\rho}{8} \partial^2}\Psi ^2 \right)\overleftrightarrow
{\partial}\left(\partial e^{\frac{\rho}{8}{\partial}^2}
\Upsilon\right),
\end{eqnarray}
here and below $\partial =\frac{d }{dt}$ and $A\overleftrightarrow
{\partial}B= A\partial B - (\partial A) B $. This expression is
analogous to an expression for the energy obtained for p-adic
equation from calculation of the energy-momentum tensor
\cite{0207107,Yang}.

One can immediately see that this energy conserves. Indeed,
$$
\frac{d E(t)}{dt}=- \frac{1}{2} \upsilon \partial
\upsilon+q^2(\partial \psi)(\partial ^2 \psi)-\psi
\partial \psi +\frac{1}{2}(\partial
{\Upsilon }){\Psi}^2 +{\Upsilon }{\Psi}\partial {\Psi}
$$
$$
+\frac{1}{8} \int_0^1 d\rho( e^{-\frac{\rho}{8} {\partial}^2}
{\Upsilon {\Psi}})\overleftrightarrow{\partial^2 }({\partial}
e^{\frac{\rho}{8}{\partial}^2} \Psi) +\frac{ 1 }{16}\int_0^1
d\rho( e^{-\frac{\rho}{8} {\partial}^2} {
\Psi}^2)\overleftrightarrow{\partial^2}(
e^{\frac{\rho}{8}{\partial}^2}\partial  \Upsilon).
$$
Using the  identity
\begin{eqnarray}
\label{idd} -m\int\limits_0^1 d\rho (e^{ -m\rho
\partial^2 }\varphi ) \overleftrightarrow{\partial^2  } (e^{ -m(1
- \rho)\partial^2 } \phi)= \varphi \overleftrightarrow { e^{ -m
\partial^2 } }  \phi ,
\end{eqnarray}
we get
$$
\frac{d E(t)}{dt}=- \frac{1}{2} \upsilon \partial
\upsilon+q^2(\partial \psi)(\partial ^2 \psi)-\psi
\partial \psi +\frac{1}{2}(\partial
{\Upsilon }){\Psi}^2 +{\Upsilon }{\Psi}\partial {\Psi}
$$
$$
-\Upsilon \Psi \overleftrightarrow { e^{ -\frac{1}{8}
\partial^2 } }  \partial \psi -\frac{1}{2}\Psi^2 \overleftrightarrow { e^{
-\frac{1}{8}
\partial^2 } } \partial \upsilon
$$
$$
=\frac{1}{2}\partial \upsilon[-\upsilon+e^{ -\frac{1}{8}
\partial^2 }  (\Psi^2)]+\partial \psi[(q^2 \partial^2 -1)\psi
+e^{ -\frac{1}{8}
\partial^2 }  (\Upsilon \Psi)]=0.
$$

On  solutions to equation of motion the nonlocal term can be
presented as the sum of the following
 terms
\begin{equation}
\label{parts-of-energy}
 E_{nl}(t)=E_{nl1}(t)+E_{nl2}(t)+E_{nl3}(t)+E_{nl4}(t),
\end{equation}
where
\begin{subequations}
\label{4nl}
\begin{eqnarray}
\label{nl1} E_{nl1}(t)&=& \frac18\int_0^1 d\rho\left(
e^{\frac{1-\rho}{8}{\partial}^2}(-q^2 {\partial}^2+1) \psi\right)
  \left(  e^{-\frac{1-\rho}{8}{\partial}^2} \partial^2\psi
\right),\\
\label{nl2} E_{nl2}(t)&=&- \frac18\int_0^1 d\rho\left(
e^{\frac{1-\rho}{8}{\partial}^2}(-q^2 {\partial}^2+1)\partial
\psi\right)
  \left( e^{-\frac{1-\rho}{8}{\partial}^2}\partial  \psi
\right),\\
\label{nl3} E_{nl3}(t)&=& \frac{1}{16} \int_0^1 d\rho\left(
e^{\frac{1-\rho}{8} \partial^2}\upsilon \right) \left(
e^{-\frac{1-\rho}{8}{\partial}^2}
\partial^2\upsilon\right),
\\
\label{nl4} E_{nl4}(t)&=& - \frac{1}{16} \int_0^1 d\rho\left(
e^{\frac{1-\rho}{8} \partial^2}\partial \upsilon
\right)\left(e^{-\frac{1-\rho}{8}{\partial}^2}
\partial \upsilon\right).
\end{eqnarray}
\end{subequations}

It is interesting to compare the expression
(\ref{parts-of-energy}) for the nonlocal part of energy with the
nonlocal term corresponding to the energy   for  equation
(\ref{Psi-ap1}). The latter has the form (we use calligraphic
letters for quantities defined on solutions to equation
(\ref{Psi-ap1}))
\begin{equation}
\label{calE} {\cal E}(t)={\cal E}_{p}+{\cal E}_{k}+ {\cal
E}_{nl}(t),
\end{equation}
where
\begin{equation}
\label{ekep} {\cal E}_{k}=\frac{q^2}{2} (\partial \psi)^2 ,~~~~
{\cal E}_{p}=- \frac{1}{2}\psi^2+ \frac{1}{ 4}\Psi ^4,
\end{equation}
and  the nonlocal term mentioned above is given by
\begin{equation}
\label{enl} {\cal E}_{nl}(t)=\frac18\int_0^1 d\rho\left(
e^{\frac{-\rho}{8}{\partial}^2} \Psi ^3 \right)
\overleftrightarrow{\partial}
  \left(\partial  e^{\frac{\rho}{8}{\partial}^2} \Psi \right).
\end{equation}
Here $\Psi =\Psi _{app}$ and $\psi =\psi _{app}~$.
This nonlocal term takes the following form on equation of motion
\begin{equation}
\label{Enla} {\cal E}_{nl}(t)={\cal E}_{nl1}(t)+{\cal E}_{nl2}(t)=
\frac18\int_0^1 d\rho\left( e^{\frac{1-\rho}{8}{\partial}^2}(-q^2
{\partial}^2+1 ) \psi \right) \overleftrightarrow{\partial}
  \left(\partial  e^{-\frac{1-\rho}{8}{\partial}^2} \psi \right).
\end{equation}
Comparing (\ref{Enla}) to (\ref{nl1}) and (\ref{nl2}) we see that
functional expressions for ${\cal E}_{nl1}$ and ${\cal E}_{nl2}$
are the same as the ones to $E_{nl1}$ and $E_{nl2}$.

\begin{figure}[!h]
\centering
\includegraphics[width=7cm]{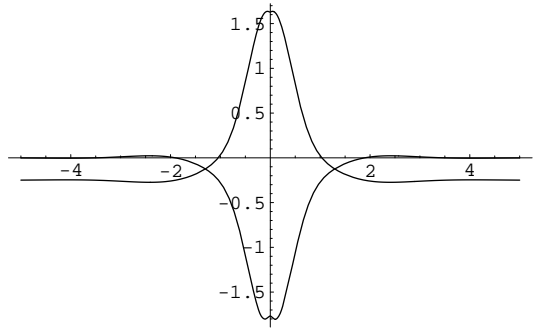}
\caption{Testing  the constancy of the energy for the
interpolating solution to equation (\ref{Psi-ap2}). This figure
shows ${\cal E}_{p}(t)+{\cal E}_{k}(t)$ and ${\cal E}_{nl}(t)$,
calculated from (\ref{ekep}) and (\ref{eint}), respectively. Their
sum is equal to zero up to numerical errors.} \label{fig:ek-p}
\end{figure}

In Fig.\ref{fig:ek-p} we plot functions ${\cal E}_{p}(t)+{\cal
E}_{k}(t)$ and  ${\cal E}_{nl}(t)$.
 We see that up to inaccuracy in numerical calculations
the energy is conserved. To calculate the nonlinear terms
numerically we actually use another representation for ${\cal
E}_{nl}(t)$,
\begin{equation}
\label{sr} {\cal E}_{nl}(t)=\frac18\int_0^1 d\rho\left(
e^{\frac{2-\rho}{8}{\partial}^2}(-q^2 {\partial}^2+1 ) \Psi
\right) \overleftrightarrow{\partial}
  \left(\partial  e^{\frac{\rho}{8}{\partial}^2} \Psi \right).
\end{equation}
(\ref{sr}) allows us to use the integral form
\begin{equation}
\label{eint} {\cal E}_{nl}(t)=\frac{1}{8}\int_{\varepsilon}^{1}
d\rho (K_1 \Psi)(K_2 \Psi)- \frac{1}{8}\int_{\varepsilon}^{1}
d\rho (K_3 \Psi)(K_4 \Psi),
\end{equation}
where $K_1=(-q^2 {\partial}^2+1)C_{\frac{2-\rho}{8}},
~~K_2={\partial}^2C_{\frac{\rho}{8}}, ~~K_3=(-q^2
{\partial}^3+\partial)C_{\frac{2-\rho}{8}},
~~K_4={\partial}C_{\frac{\rho}{8}}.$ In  actual calculations we
take $\varepsilon \simeq 0.02$.

\subsection{Pressure}
The pressure is defined in terms of the energy-momentum tensor
$P(t)_i=-T_i^i$ (no summation). Since we consider homogeneous
configurations we will omit the vector index $i$. The pressure has
the form
\begin{equation}
\label{pressure} P(t)=-E+2E_{k}(t)+2E_{nl2}(t)+2E_{nl4}(t),
\end{equation}
or explicitly
\begin{equation}
\label{pres}
\begin{split}
P(t)=-E+q^2 (\partial  \psi)^2&-\frac14 \int_0^1 d\rho(
e^{\frac{1-\rho}{8}{\partial}^2}( -q^2{\partial}^2+1)\partial\psi)
  ( e^{-\frac{1-\rho}{8}{\partial}^2}\partial  \psi )\\
&-\frac18 \int_0^1 d\rho( e^{\frac{1-\rho}{8}
{\partial}^2}\partial \upsilon  )(
e^{-\frac{1-\rho}{8}{\partial}^2} \partial\upsilon).
\end{split}
\end{equation}

The analogous formula takes place for bosonic SFT
\cite{0207107,Yang}. For equation (\ref{Psi-ap2}) the pressure
${\cal P}$ is
\begin{equation}
\label{pressure-1} {\cal P}(t)=
  -E+2{\cal E}_k(t)+2{\cal E}_{nl2}(t),
\end{equation}
i.e. the functional expression for this pressure is
 given by
\begin{equation}
\label{p1} {\cal P}(t)=-E +q^2 (\partial  {\cal \psi})^2-\frac14
\int_0^1 d\rho( e^{\frac{1-\rho}{8}{\partial}^2}(
-q^2{\partial}^2+1)\partial_{t}\psi)
  ( e^{-\frac{1-\rho}{8}{\partial}^2}\partial_{t}  \psi ).
\end{equation}
For a large time the pressure in both cases is equal to $-E$.

One can see that  the pressure (\ref{p1}) at the zero  time is
equal to the total energy $E$, i.e. ${\cal P}(0)=E.$ This can be
shown by taking into account the explicit relation between the
energy and the pressure (\ref{pressure-1}) and the fact that for
our solution ${\cal E}_{nl1}(0)=0$. Indeed,
$$
{\cal P}(0)=-E+q^2 (\partial  \psi)^2(0)+{\cal E}_{nl2}(0)= -E+q^2
(\partial  \psi)^2(0)+2E-q^2 (\partial  \psi)^2(0)=E.
$$
For the case of system of equations (\ref{seq}) the pressure at
$t=0$ is given by
\begin{equation}
\label{p0-se} P(0)=-E+q^2 (\partial \psi)^2(0)-\frac14 \int_0^1
d\rho\left( e^{\frac{1-\rho}{8}{\partial}^2}(
-q^2{\partial}^2+1)\partial\psi\right)
  \left( e^{-\frac{1-\rho}{8}{\partial}^2}\partial  \psi \right).
\end{equation}
One sees that this functional expression coincides with the
expression (\ref{p1}) that gives the pressure for
 equation (\ref{Psi-ap2}).
\begin{figure}[!h]
\includegraphics[width=7cm]{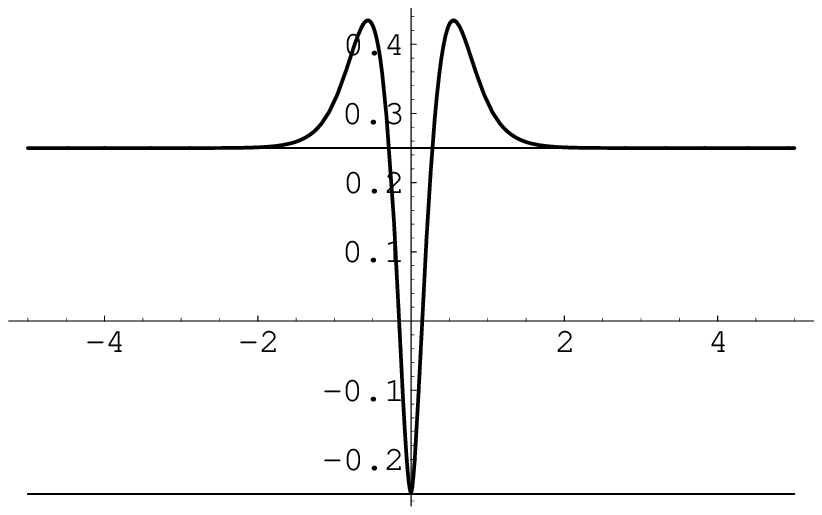}a $~~~$
\includegraphics[width=7cm]{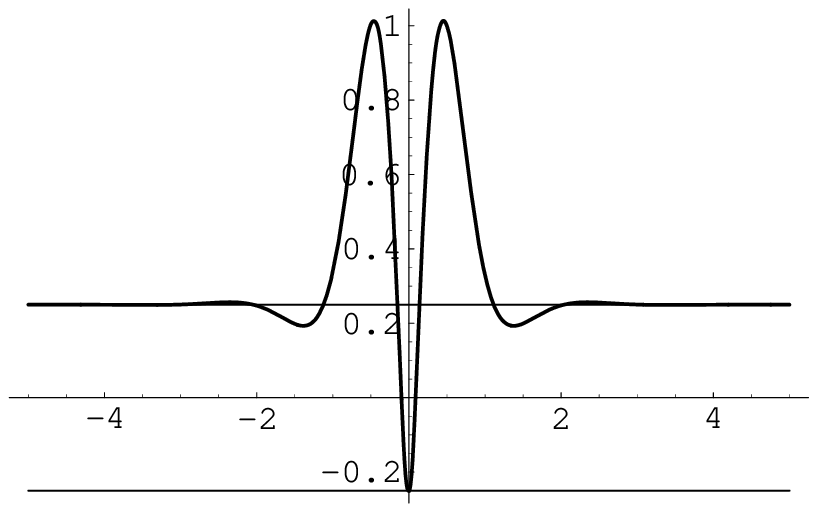}b
\centering \caption{Calculating the pressure on the solutions
Fig.\ref{f12} to  equation (\ref{Psi-ap2}); a) for $q=0$, $~~~$ b) for
$q^2=0.96$. This figure shows ${\cal P}(t)$ has  minimum at $t=0$
and goes to 1/4 at the $\pm \infty$  } \label{Fig-press}
\end{figure}
In Fig.\ref{Fig-press} we show the pressure ${\cal P}(t)$ for
equation (\ref{Psi-ap2}) as a function of time. To obtain this
picture we use
\begin{equation}
{\cal P}(t)=-E+q^2 (\partial  \psi)^2(0)-
\frac14\int_{\varepsilon}^{1} d\rho (K_3 \Psi)(K_4 \Psi).
\end{equation}
The minimum of the pressure is  at t=0. Note that for  large times
the value  of ${\cal P}$ is positive and  equal to  $-E=1/4$.

In the above consideration we take into account only the energy of
tachyon field. Following  the Sen conjecture one has to add to the
action an extra term representing the D-brane tension. This term
shifts on a constant the tachyon potential making it  equal to
zero at its  minimum. This constant also shift the pressure
\cite{Yang} and we get that for our solutions the pressure goes to
zero for large times.

\section{Conclusion}
In this paper we have  studied the  SSFT tachyon classical time dependent
equations of motion
describing  non-stability of non-BPS D-branes.
We have found the special time dependent solution to this equation
that describes the rolling tachyon.
It starts from the perturbative vacuum and  approaches   one of
stable vacua  in infinite time.
We have argued that the pressure on this solution tends to zero for large times.
These results have been obtained in the first non-trivial level within the level truncation
method.
In the next paper  we will  take into account  higher
level fields including vector fields and  study a dynamics of a
D-brane tension.

\section*{Acknowledgements}

We would like to thank D.~Belov, A.~Giryavets, B.~Dragovic, N.A.~Slavnov, I.~Tyutin
and I.V.~Volovich for
fruitful discussions.  We especially thank Ya.~Volovich for providing us his numerical results.
A.K. acknowledges stimulating discussions with E.~Kiritsis.

This work is supported in part by RFBR grant 02-01-00695. I.A. and A.K. are
supported in part by RFBR grant for leading scientific schools and
by INTAS project 99-0590. A.K. is supported in part by INTAS
fellowship for Young Scientists YSF~2002-42.

\end{document}